\documentclass{article}
\usepackage{amsfonts}

\usepackage{amsmath}

\input{tcilatex}

\begin{document}

\author{Bert Schroer \\
presently: CBPF, Rua Dr. Xavier Sigaud, 22290-180 Rio de Janeiro, Brazil \\
email: schroer@cbpf.br\\
Prof. emeritus of the Institut f\"{u}r Theoretische Physik\\
FU-Berlin, Arnimallee 14, 14195 Berlin, Germany}
\title{Braided Structure in 4-dimensional conformal Quantum Field Theory \\
{\small Dedicated to Gerhard Mack and Robert Schrader on the occasion of
their 60th} {\small birthday}\\
{\small submitted to Phys. Lett. B}}
\date{November 2000}
\maketitle

\begin{abstract}
Higher dimensional conformal QFT possesses an interesting braided structure
which, different from the d=1+1 models, is restricted to the timelike region
and therefore easily escapes euclidean action methods. It lies behind the
spectrum of anomalous dimensions which may be viewed as a kind of substitute
for a missing particle interpretation in the presence of interactions.
\end{abstract}

\section{Introductory remarks}

Since the early 60ies the use of conformal field theory in particle physics
has been beset by physical doubts concerning incompatibilities with the LSZ
framework of interacting (Wigner) particles. Indeed, besides free zero mass
particles there are strictly speaking no other conformal situations which
are consistent with a particle interpretation \cite{Sch1}. The analytic
simplifications of massless limits are (again apart from free situations) in
some sense paid for by the conceptual complications within the particle
setting; the latter allow at best to extract extremely inclusive scattering
data from interacting conformal field theory using a scattering theory which
is based on probabilities instead of amplitudes \cite{Bu} i.e. conformal
theory is not directly a theory of particles. Nevertheless, as it is already
well known from chiral conformal theories, conformal field theory with its
arena of spacetime charge flows and their fusions is presently the most
useful theoretical laboratory to explore nonperturbative aspects of QFT in a
mathematically controllable way.

This mentioned particle physics weakness was more than compensated for by
the profound application to the theory of critical phenomena in statistical
mechanics. In fact there have been two approaches in that direction. The
older and in its aims more ambitious and general way is \textit{extrinsic}
i.e. one tries to approach critical theories from a generic i.e. noncritical
setting by the renormalization group action in the commutative Euclidean
functional integral setting. This is the method of \footnote{%
A formulation in the real-time setting of Wightman (i.e. noncommutative in
the old-fashioned sense) which probably requires the use of more subtle
conditional expectations on operator algebras has never been achieved. On
the other hand there are also several real-time properties whose statistical
mechanics interpretation is unknown to this date, e.g. the issue of the
timelike structure in this article.} either Wilson (momentum space) or
Kadanoff (x-space). These methods work well in the standard Euclidean
setting if the functional integral approach coupled with dimensional
perturbative ideas; it gives surprisingly good numerical results in the
Lagrangian Callan-Symanzik setting. Its main aim is to obtain a numerical
approximation for the critical indices of a given Hamiltonian or action in a
continuous or discretized system.

The advent of chiral conformal models exemplified (in a more restricted
setting) another, but this time truely intrinsic approach. In this the
critical theories where not constructed by finding renormalization group
transformations in an ambient coupling constant space which drive the system
towards the scale invariant fixed points, but rather by finding enough
intrinsic properties of conformal models which eventually lead to their
classification and explicit constructions. Since the detailed connection
between the concrete condensed matter and its microscopic description is in
many cases unknown anyhow, this second method, if combined with the
universality class concept and precise measurements really explains the
critical indices as anomalous dimensions in an associated real time
conformal field theory. As we will see below, they are related to a
classifiable and computable algebraic structure. A more detailed application
to a classification of exact critical indices of systems related to higher
dimensional conformal field theories will be presented elsewhere \cite{S}.

A prototype of such an intrinsic approach is the BPZ scheme \cite{B-P-Z} of
conformal blocks\footnote{%
It was foreshadowed by Kadanoff's ideas on ``Coulomb Gas Representations''.}
which was successfully applied towards a construction of minimal models.
There one aims at the critical theories right away and worries about their
noncritical deformations (which convert them into islands in a continuum of
theories in one phase) afterwards. The mathematical basis are certain
algebras (Virasoro-, current-,W-) and their positive energy representations.

Unfortunately these algebras are limited to two-dimensional QFT, but
fortunately it turned out that one can understand a substantial part of the
two-dimensional situation by adapting the general theory of superselection
rules of Doplicher-Haag-Roberts to low dimensions \cite{Haag}, thus
increasing the chances of generalizations away from two dimensions. Whereas
the Wilson-Kadanoff method had no counterpart on the noncommutative real
time side, the real-time DHR method has no known Euclidean formulation. The
power of this more noncommutative DHR method of superselection charges in
chiral theories resembles somewhat the power of the more noncommutative
transfer matrix method in Baxter's work on 2-dim. lattice models in
statistical mechanics. As a result of the restrictive power of relativistic
causality and locality and the ensuing vacuum structure which distinguishes
QFT from quantum mechanics, the method of algebraic QFT (AQFT) is more
generic than the transfer-matrix method. In particular its validity is not
limited to subclasses of chiral theories, but promises to classify and
construct all of them \cite{FRSII}.

The main point is the classification of admissible braid-group statistics 
\cite{Rehren}. This is part of the classification of superselected charges
and their fusion. It does not require the study of genuine space-time
properties of the dynamics of charge flows and can be done in terms of
combinatorial algebras (type II$_{1}$ von Neumann algebras, in particular
braid group intertwiner algebras with Markov traces on them). This is
analogous (but more difficult) to the internal symmetry data for free field
theory. As in the latter case the spin/statistics+group representation data
uniquely fix the free field, the spacetime aspects of chiral models are
completely determined by the charge superselection data, albeit in a quite
sophisticated way. This tight connection continues to hold in chiral theories%
\footnote{%
Whenever there is a continuously varying coupling parameter as in the
massless Thirring model, this can be absorbed into the chiral statistics
parameters (= chiral anomalous dimensions modulo integers).} but perhaps
canot be expected for higher dimensional conformal theories; the present
investigation does not try to resolve the problem whether in addition to the
time- and spacelike- superselection data there are also deformation
parameters which locally do not change the superselection class ($\simeq \,$%
``phases$"$ separated by phase transitions) but modify the strength of
interactions.

The restrictive role in terms of fusion and statistics of superselected
charges is already visible in the analysis of 4-point functions from
monodromy and short distance behavior in chiral conformal theories, but a
systematic construction without relying on educated guesses and consistency
checks still awaits elaboration. In any case the intrinsic approach which
aims first at the explicit constructions of the conformal ''islands'', and
then tests the infinitesimal surrounding by some kind of Zamolodchikov
perturbation idea, has been quite successful and promises more to come.

The drawback of the exact intrinsic method is that it appears limited to
chiral theories. Higher dimensional conformal field theories do not have the
simplifying feature of chiral factorization, but they also seem to be
formally close to free field theories in the aforementioned sense. As far
back as the middle 70ies there were two ways of thinking about general
conformal fields, either as Wightman fields on the covering of compactified
Minkowski space with a globalized notion of causality and the
field-state-vector relation (absence of local annihilators, i.e. the
Reeh-Schlieder property) as it appeared in the work of Luescher and Mack 
\cite{L-M}, or the (controllable) nonlocal projection back into the
compactified Minkowski spacetime as a kind of quasiperiodic operator-valued
sections. Different from (smeared) Wightman fields, the projected fields
have a source and range projector attached to them\footnote{%
In chiral theories these operators obey exchange algebra relations and are
often referred to as vertex operators as opposed to chiral observables which
are described by Wightman fields.} and arise as the components of the L-M
Wightman fields in a decomposition with respect to the center of the
conformal covering group. The resulting decomposition theory of the covering
fields appeared in joint work involving the present author \cite{S-S} and
was done independent of the Luescher-Mack work. It is best seen as a
timelike analogue of the spin-statistics theorem in that it relates the
phases under a full timelike sweep through the Dirac-Weyl compactified
Minkowski world to the spectrum of anomalous scale dimensions modulo
(half)integers (in the spin case the spatial $2\pi $ rotational sweep). This
raises the question whether there could be also some timelike ``statistics''
(exchange structure) behind the anomalous dimension spectrum. It is the
purpose of this short note to argue that this is indeed the case.

\section{The Timelike Plektonic Structure}

In a situation without controllable models one has to proceed in a
structural way and at the end use the obtained structure to classify and
construct models. As mentioned in the introduction there are two ways to
explore the structure of conformal theories, the Luescher-Mack approach
which uses Wightman fields on the conformal covering space and the approach
of Swieca and myself which works with nonlocal quasiperiodic component
fields on the compactified ordinary Minkowski spacetime. The connection is
given by the decomposition formula for conformal fields

\begin{align}
F(x)& =\sum_{\alpha ,\beta }F_{\alpha ,\beta }(x),\,\,F_{\alpha ,\beta
}(x)\equiv P_{\alpha }F(x)P_{\beta }  \label{dec} \\
Z& =\sum_{\alpha }e^{2\pi i\theta _{\alpha }}P_{\alpha }  \notag
\end{align}
Here $Z$ is the generator of the center of the conformal covering group $%
\widetilde{SO(4,2)}$, $P_{\alpha }$ are its spectral projectors and the
phases are the mod1 reductions of the anomalous dimensions associated with
the family of fields sharing the same phase (conformal blocks). Whereas the $%
F(x)$ are L\"{u}scher-Mack fields i.e. globally causal Wightman fields which
live on the covering space (and therefore one should use the appropriate
more complicated coordinates or charts), the $F_{\alpha \beta }(x)$ are
(operator-valued) trivializing section on the Dirac-Weyl compactification $%
\bar{M}$ are apart from fulfilling a timelike numerical quasiperiodicity
condition conventional objects of the ``laboratory-world'' (without
requiring the heaven and hells of the covering) \cite{S-S-V}\cite{K-R-Y}. In
chiral theories they become the ``vertex operators'' appearing in the
conformal block analysis of \cite{B-P-Z}.

Although in higher dimension one cannot have anything else than spacelike
Bose/Fermi commutation relation of the standard field/particle statistics,
the possibility of defining conformal observables as local fields which live
on the compactification $\bar{M}$ prepares the ground for becoming aware of
timelike commutativity (Huygens principle). Together with the ordering
structure inside the timelike lightcone this allows for a consistent
timelike braid group commutation structure. In terms of the above double
indexed fields this means the validity of the following exchange algebra

\begin{align}
F_{\alpha ,\beta }(x)G_{\beta ,\gamma }(y)& =\sum_{\beta ^{\prime }}R_{\beta
,\beta ^{\prime }}^{(\alpha ,\gamma )}G_{\alpha ,\beta ^{\prime
}}(y)F_{\beta ^{\prime },\gamma }(x),\,\,x>y  \label{time} \\
F_{a,\beta }G_{\beta ,\gamma }& =\sum_{\beta ^{\prime }}R_{\beta ,\beta
^{\prime }}^{(\alpha ,\gamma )}G_{\alpha ,\beta ^{\prime }}F_{\beta ^{\prime
},\gamma },\,\,locF>locG  \label{ex}
\end{align}
where the second line refers to not necessarily pointlike localized
operators. The arguments that the R-matrices must be representers of the
Artin braid group is the same as in d=1+1 where the spacelike region has the
same topology as timelike light cones. These component fields have no
physical role in the spacelike region in fact they are local and one has to
sum them according to (\ref{dec}) in order to formulate Einstein causality.
This timelike braid structure is of course consistent with the structure of
the conformal 2- and 3-point functions. For example from

\begin{align}
\left\langle F(x)F(y)^{\ast }\right\rangle & \simeq lim_{\varepsilon
\rightarrow 0}\frac{1}{\left[ -\left( x-y\right) _{\varepsilon }^{2}\right]
^{\delta _{A}}}\, \\
\left( x-y\right) _{\varepsilon }^{2}& =\left( x-y\right) ^{2}+i\varepsilon
(x_{0}-y_{0})  \notag
\end{align}
one reads off the timelike relation 
\begin{equation*}
\left\langle F(x)F(y)^{\ast }\right\rangle =e^{2i\delta _{F}}\left\langle
F(y)^{\ast }F(x)\right\rangle \,,\ x>y
\end{equation*}

Let us now look at this situation from a more general analytic viewpoint of
vacuum correlation functions. It is well known that Wightman functions allow
for analytic continuations into the so-called BWH domain. As was observed
first by Jost this complex domain has real spacelike points which he was
able to describe explicitly (Jost points). This fact is useful if one wants
to describe the various operator orderings inside an n-point function in
terms of one analytic master function by using spacelike local
commutativity. What is less well known is that the use of conformal
invariance allows for an additional analytic extension a la BHW which leads
to timelike Jost points. Again the role of algebraic relations is to bind
together the various timelike orderings into one analytic master function.
But now, as a consequence of the more general braid structure one will loose
the uniqueness in that continuation i.e. one will end up with ramified
coverings which are the analytic manifestations of Huygens-localizable
fields (relative to the observables) which are timelike nonlocal (plektonic)
among themselves. Whereas the intermediate projectors nearest to the vacuum
are uniquely determined by the anomalous dimension of the field, there are
the well-known branchings in other projectors. For the 4-pointfunction we
obtain a sum of several contributions which are interrelated by crossing
transformations which correspond to the application of the plektonic
commutation relations.

\begin{align}
& W(x_{4},x_{3},x_{2},x_{1}):=\sum_{\gamma }\left\langle
F(x_{4})F(x_{3})P_{\gamma }F(x_{2})F(x_{1})\right\rangle \\
& =\left[ \frac{x_{42}^{2}x_{31}^{2}}{(x_{43})_{\varepsilon
}^{2}(x_{32})_{\varepsilon }^{2}(x_{21})_{\varepsilon
}^{2}(x_{14})_{\varepsilon }^{2}}\right] ^{\delta _{A}}\sum_{\gamma
}w_{\gamma }(u,v),\,\,  \notag \\
& \,\,\,u=\frac{x_{43}^{2}x_{21}^{2}}{(x_{42})_{\varepsilon
}^{2}(x_{31})_{\varepsilon }^{2}},\,\,v=\frac{x_{32}^{2}x_{41}^{2}}{%
(x_{42})_{\varepsilon }^{2}(x_{31})_{\varepsilon }^{2}}  \notag
\end{align}
This is all very similar to the well studied chiral situation except that
the analytic consequences of monodromy relations are somewhat more involved
because there is no factorization which reduces the number of cross ratios.
The new aspect as compared to the exchange structure of chiral theories is
the issue of compatibility between the plektonic timelike- anf the spacelike
bosonic/fermionic exchange algebra. This issue was discussed in the above
analytic setting of correlation functions in \cite{Huy}. From these
consistency relations Rehren \cite{Re2} abstracted the following group
structure which amalgamates the Artin braid group $B_{\infty }$ with the
symmetry group $S_{\infty }$ (``mixed group'' $G_{\infty })$

\begin{eqnarray}
b_{i}t_{j} &=&t_{j}b_{i},\,\,\left| i-j\right| \geq 2 \\
b_{i}t_{j}t_{i} &=&t_{j}t_{i}b_{j},\,\left| i-j\right| =1  \notag \\
b_{i}b_{j}t_{i} &=&t_{j}b_{i}b_{j},\,\,\left| i-j\right| =1\,\,  \notag
\end{eqnarray}
Here the b's are the generators of the braid group, i.e. they fulfill the
well-known Artin relations among themselves, whereas the t's are the
transpositions which generate the permutation group.

The question of how these consistency problems can be taken care of \ within
the conceptually more complete and mathematically more rigorous DHR setting 
\cite{Haag} still awaits solution; the problem here is that the spacelike
Haag dualization generates a different net structure from the Haag dual net
on $\bar{M}.$ Whereas this is welcome since one wants to arrive at two
different localized endomorphisms and this is only possible on two different
nets, the exact positions of these nets remain an open problem for further
research.

One important aspect of the endomorphism formalism would be the
understanding of the central charges which are expected to arise from the
global selfintertwiners related to the timelike charge transport around $%
\bar{M}$ and their associated S-T $SL(2,\mathbb{Z})$-structure \cite{FRSII}.
Whereas one expects the physical realization in the sense of Rehren's
``statistics characters'' \cite{Re-Pal} to continue to hold in higher
dimensions, the derivation of  modular identities based on Verlinde's
geometric argument which led to the modular ``duality'' (dual temperature $%
\beta \rightarrow \frac{2\pi }{\beta }$) breaks down. This can be seen by
explicitly calculating the partition function of \ the conformal Hamiltonian 
$R_{0}$ \cite{Huy} which is equal to the true Hamiltonian of the associated
anti- deSitter model. In the case of $AdS_{4}$ for which Fronsdal \cite
{Frons} computed its spectrum (i.e. $R_{0}$ in 3-dim. conformal theory),one
obtains 
\begin{eqnarray}
tre^{-\beta R_{0}} &=&e^{-E_{0}\beta }\prod_{n=1}^{\infty }\frac{1}{\left(
1-q^{n}\right) ^{N_{n}}},\;\;N_{n}=2n+1 \\
R_{0} &=&E_{0}+\sum_{n=1}^{\infty }\sum_{k=1}^{\left( 2n+1\right)
}na_{n}^{(k)\ast }a_{n}^{(k)}  \notag
\end{eqnarray}
and it is clear that this partition function does not have the properties of
the $\eta $-and Jacobi $\vartheta $-functions which appear in the chiral
theories. In other words one needs the chiral factorization in order to
validitate Verlinde's differential geometric torus interpretations, i.e.
causality and braid group aspects alone are not sufficient to produce
modular $SL(2,R)$ relations in thermal partition functions of $R_{0}.$

\section{Structural Discoveries in LQP and Murphy's Law}

There are three ways in which Local Quantum Physics reacts to inventions of
particle physicists. On the one hand there are those ideas which although
not required by the general principles are nevertheless consistent because
they allow for a model realization (perturbatively, or low-dimensional
nonperturbatively). A good example is supersymmetry; in the absence of
experimental indications the theoretical ``naturality'' of such an object
may depend on what aspect of the formalism of QFT one considers as
physically important. The majority of inventions in QFT, especially those
which tried to modify QFT by cutoffs, nonlocalities etc. suffered a less
lucky fate and were in most cases doomed by Murphy's law \cite{Mur} (``what
can go wrong, will go wrong'') which asserts its strength in a rather swift
way especially in the precision setting of local quantum physics.

On the opposite end are the attempts to reformulate the content of quantum
field theory in a intrinsic field-coordinate independent way, using methods
of operator algebras and aiming not at inventions with unclear relations to
the physical principles, but rather unraveling structural properties as
consequences of the physical principles (which may have escaped the standard
quantization formalism). The present observation about a timelike algebraic
structure in conformal theories may serve as an illustration of a structural
property which is not imposed from the outside and hard to be seen by
Lagrangian methods. Rather it is derived from requiring a local explanation
for the dynamical timelike superselection structure which in turn is
responsible for the appearance of anomalous dimensions. It has been a
remarkably gratifying experience that all admissible structures which had
been derived by classifying the modes of realizations of the underlying
principles sooner or later were realized through the construction of new
models. For example in the case of the lightlike braid group structure of
chiral models all serious attempts to find a model realization of admissible
charge superselection with given fusions and braid group commutation
relations (exchange algebras) were successful and lend weight to the
validity of an ``anti-Murphy'' maxim: \textit{every structure which is
derived from the general principles of local quantum physics has a model
realization.}

The specific connection between the anomalous dimensional spectrum and the
plektonic data not only depend on spacetime dimensions but for a given
spacetime dimension they also depend on the nature of the spacelike
superselections e.g. on the presence of group representation multiplicities;
e.g. charge-carrying fields in current algebras have a different dimensional
spectrum than W-algebras even if the fusion laws and the braid group
statistics is the same. The dynamic aspect of timelike superselections
nourishes the hope that there are no coupling parameters which cannot be
encoded into the spectrum of anomalous dimensions\footnote{%
This is the situation in chiral theories where e.g.the coupling strength of
the massless Thirring model may be completely absorbed into the variable
anomalous dimensions of its chiral components.
\par
{}}. This would uniquely relate the spacetime nets with the superselection
structure (similar to the case of chiral theories) and significantly
facilitate their actual construction.

It should be clear at this point that the main purpose of our interest in
conformal theories is the actual construction of a nontrivial 4-dimensional
field theory in order to show that the principles of 20 century QFT are
sound and allow a mathematically consistent presentation as all previous
areas in physics. Principles as causality/locality should be limited and
superseded by new principles and not by an ad hoc cut-off in certain
Feynman- or functional- integrals and the conceptual havoc created by
cutoffs is most evident in conformal theories. 

The standard perturbation theory is not directly applicable to zero mass
situations\footnote{%
Note that interacting conformal theories are not theories of particles since
a massless particle has necessarily an interpolating field with
(``protected'') canonical dimension and that in turn leads to an
interaction-free sector, i.e. all fields whose applications to the vacuum do
not lead out of this sector (which includes in particular all composites of
the interpolating field) are also void of interactions \cite{Sch1}.} (e.g.
the infrared divergencies of massless QED); the strategy to obtain
perturbative conformal models has been to look out for massive models with
vanishing Beta-functions ($\simeq $vanishing of coupling constant
renormalization) which is certainly necessary for conformal invariance but
also seems to create that ``soft'' mass dependence which guaranties the
existence of the massless limit. Although every conformal theory is
automatically renormalizable in the sense of an order-independent dominating
short distance power law, the Lagrangian candidates for conformal field
theories are extremely scarce; the only known candidates in d=1+1 are very
special 4-Fermion couplings as the Thirring model whereas for d=1+3 one
seems to be restricted to a particular supersymmetric Yang-Mills models
(SYM) namely those with vanishing beta-functions. Most chiral models have no
coupling strength which allows to deform them continuously into free
theories; rather such coupling constants are only expected to arise in a
Zamolodchikov perturbation around the conformal islands; there is no reason
to expect an improvement of this situation in d=1+3, which means that the
above scarcity would be a perturbative fake.  

At this point one may think that the much discussed AdS-CQFT connection
could enrich the d=1+3 conformal situation by applying the perturbation on
the AdS side where there is no (Beta-function) restriction besides the
SO(4,2) invariance. Indeed the rigorously proven isomorphism\footnote{%
In the formulation of the Maldacena-Witten conjecture it is tacitely assumed
that to one conformal theory there corresponds exactly one AdS theory but
without leaving the realm of pointlike fields it is not possible to prove
this.} \cite{ReI} does secure the existence of a conformal theory but
unfortunately the holographic images of all AdS pontlike field theories are
conformal theories which violate the causal propagation (causal shadow)
property since the pointlike AdS theory contains too many degrees of freedom
for a physically acceptable conformal theory \cite{Huy}. On the other hand a
conformal theory with the correct causal propagation leads to a AdS theory
in which the best localized generators are ``strands'' i.e. stringlike
objects which however do not increase the conformal degrees of freedom as
the dynamical strings of string theory would do. 

I believe that this new timelike structure will just leave enough space to
sail between the scilla of trivial free field theories (as expressed by the
triviality of an imposed interacting particle structure) and the charybdis
of mathematically uncontrollable ``normal'' quantum field theories. In this
way I expect 4-dimensional conformal field theories to be the first
nontrivial nonperturbatively controllable and explicitly constructed QFT in
physical spacetime. It would be nice if finally, after almost 50 years of
strenuous efforts, local quantum field theory could enjoy the same status as
any other mathematically and conceptually established physical theory.

\end{document}